\documentclass[prl,twocolumn,showpacs,preprintnumbers,amsmath,amssymb,floatfix]{revtex4-1}
\usepackage{graphicx}
\usepackage{dcolumn}
\usepackage{bm}
\usepackage{color}

\begin{document}

\title{The importance of intra-molecular electron spin relaxation in small molecular semiconductors}

\author{L. Schulz$^{1}$, M. Willis$^{2}$, L. Nuccio$^{2}$, P. Shusharov$^{2}$, S. Fratini$^{3}$, F. L. Pratt$^{4}$, W. P. Gillin$^{2}$, T. Kreouzis$^{2}$, M. Heeney$^{5}$, N. Stingelin$^{5}$, C. A. Stafford$^{6}$, D. J. Beesley$^{5}$, C. Bernhard$^{1}$, J.E. Anthony$^{7}$, I. Mckenzie$^{4}$, J. S. Lord$^{4}$, A. J. Drew$^{1,2}$}

\affiliation{${}^1$D\'epartement de Physique and FriMat, Universit\'e de Fribourg, CH-1700 Fribourg, Switzerland}
\affiliation{${}^2$ Queen Mary University of London, Department of Physics, London, E1 4NS, UK}
\affiliation{${}^3$Institut N\'eel CNRS, F-38042 Grenoble, FR}
\affiliation{${}^4$ISIS Muon Facility, Rutherford Appleton Laboratory, Didcot, OX11 0QX, UK}
\affiliation{${}^5$Centre for Plastic Electronics, Imperial College London, SW7 2AZ, London, UK}
\affiliation{${}^6$Department of Physics, University of Arizona, Tucson, USA}
\affiliation{${}^7$Department of Chemistry, University of Kentucky, Lexington, KY, 40506 USA}

\date{\today}

\begin{abstract} 
Electron spin relaxation rate (eSR) is investigated on several organic semiconductors of different morphologies and molecular structures, using avoided level crossing muon spectroscopy as a local spin probe. We find that two functionalized acenes (polycrystalline tri(isopropyl)silyl-pentacene and amorphous 5,6,11,12-tetraphenyltetracene) exhibit eSRs with an Arrhenius-like temperature dependence, each with two characteristic energy scales similar to those expected from vibrations. Polycrystalline tris(8-hydroxyquinolate)gallium shows a similar behavior. The observed eSR for these molecules is no greater than 0.85 MHz at 300 K. The variety of crystal structures and transport regimes that these molecules possess, as well as the local nature of the probe, strongly suggest an intra-molecular phenomenon general to many organic semiconductors, contrasting the commonly assumed spin relaxation models based on inter-molecular charge carrier transport.

\end{abstract}

\maketitle
Research into organic semiconductors has progressed at a phenomenal rate over the last few decades, with great success in moving from the initial stages of understanding the fundamental physics, through to today where products are entering the market \cite{Berggren}. Besides their ability to transport charge or emit light, there has recently been an interest in organic semiconductors as  candidates for future spin-based technologies. This is thanks to their long spin coherence time and their suitability as a system for studying the fundamental spin phenomena relevant to many materials and applications \cite{Dediu1,Drew,Cinchetti, Szulczewski}. 

Traditional explanations for their long spin coherence time, $\tau_e$, are a weak hyperfine coupling of the electronÕs spin to the nuclear spins  \cite{Pramanik} or a weak spinÐorbit (SO) interaction due to organic semiconductors being composed of light species \cite{Rybicki}. Currently, it is not clear what the dominant intrinsic relaxation mechanism is, as the experimental measurements of $\tau_e$ vary between a few microseconds to a second or more \cite{Cinchetti, Dediu2} and there is a lack of experimental techniques that can probe $\tau_e$ microscopically. 
Electron paramagnetic resonance (EPR) is potentially capable of providing this information, but it requires a material with intrinsic electrons or electrons provided by doping, which is not always possible. 

Spin coherence times can also be extracted from magnetotransport measurements on unipolar spin valve devices \cite{Dediu1, Pramanik, Dediu2}. However, this approach requires a robust theoretical model; those applied thus far are analogues of the Elliott-Yafet \cite{Elliott} or DÕyakonov-Perel  mechanisms \cite{Dyakonov}, or one involving electrons observing a random hyperfine field upon hopping between molecular sites \cite{Bobbert}. All of these theoretical models are transport based and require an accurate measure of the charge carrier mobility, whose uncertainty \cite{Szulczewski,Park1} even in well-defined OLED structures leads to a significant error in the value of $\tau_e$. In the following, we show that muon spin relaxation/resonance ($\mu$SR) can be used to measure the intrinsic electron spin relaxation rate and we find strong evidence for an intra-molecular based spin relaxation mechanism in several different molecules with an acene backbone. Furthermore, from measurements on a rather different molecule based on hydroxyquinolate, it appears that this mechanism may be quite general.

\begin{figure}
\includegraphics[scale=0.75]{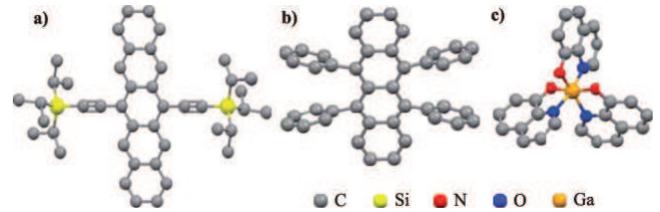}
\caption{(Color online) Structure of a) TIPS-pentacene, b) rubrene and c) Gaq3 \cite{Anthony2, Anthony3, Jurchescu, Brinkmann}.}
\label{fig1}
\end{figure}

Positively charged surface muons, with near 100 \% spin polarization, implanted into organic materials with unsaturated bonds can either thermalize as diamagnetic positively charged species or form a hydrogen-like species, known as muonium \cite{Roduner1, Pratt1, Patterson, Drew2}, that can react with a molecule creating a paramagnetic molecule. In a  "longitudinal field" experiment a magnetic field is applied parallel to the muonÕs initial spin direction, and the energy levels of the singlet/triplet bound muonium states are tuned via the Zeeman interaction \cite{Patterson}. In high magnetic fields, where most experiments are carried out, the eigenstates of the spin systems are to a good approximation pure Zeeman product states. Muons injected with their spins parallel or anti-parallel to the field are thus in an eigenstate, and no time evolution of spin polarization is expected \cite{Keitzman}. At a particular longitudinal field cross relaxation effects produce an avoided level crossing (ALC), at what would otherwise be energy level degeneracies, and eigenstates are mixtures between two Zeeman states. This leads to an oscillation between the levels. Whenever the two mixing levels belong to different muon magnetic quantum states, the time evolution causes a depolarization of the muon. The positions and linewidths of these so-called ALC resonances are determined by the muon-electron isotropic and anisotropic hyperfine coupling (HFC) \cite{Patterson}. If the hyperfine interaction is isotropic on the time scale of the experiment (normally associated with a liquid), there are only muon-nuclear spin flip-flop transitions driving the ALC resonance \cite{Keitzman}. In solids, however, the dipolar part of the hyperfine interaction drives two additional types of resonances, a single particle muon spin flip and a muon-nuclear spin flip-flip transition. The latter is very weak, whereas the former is the most intense ALC in solid media and is an extraordinarily sensitive probe of radical dynamics \cite{Keitzman, Roduner2, Heming1}, including the electron spin relaxation rate \cite{Keitzman, Heming2}.

We initially performed ALC measurements on two functionalized acenes, based on a pentacene and tetracene backbone, shown in Figures \ref{fig1}a-b. Sublimed grade 5,6,11,12-tetraphenyltetracene (rubrene) was purchased from Aldrich. Amorphous rubrene was then prepared by first heating the powder to above its melting temperature (355$^\circ$C) between two glass slides to prevent sublimation of the material, with subsequent quenching on metal plates kept at 0$^\circ$C. Tri(isopropyl)silylpentacene (TIPS-pentacene)  was synthesized  according to a published procedure \cite{Anthony1, Anthony3}, and was additionally purified by repeated recrystallization from dichloromethane/ethanol. The measured TIPS-pentacene sample was a polycrystalline powder. These materials were chosen due to the pronounced differences in solid state packing, and as a consequence in their spin transport. TIPS-pentacene crystallizes in a 2 dimensional ÔbrickworkÕ arrangement, with close $\pi-\pi$ overlap, which results in very high charge carrier mobilities \cite{Park2, Kim}. In contrast, the rubrene investigated here was an amorphous powder and as a consequence had very poor charge transport.

The muon experiments were performed on the ARGUS, EMU and HIFI spectrometers at the ISIS Muon Source, Rutherford Appleton Laboratory and the DOLLY spectrometer at the Laboratory for Muon Spectroscopy, Paul Scherrer Institute. 115 and 190 mg of the two respective organic powders were placed into 15$\times$25 mm envelopes made from 25 $\mu$m thick silver foil (99.99 \% pure). To ensure the muons stopped in the organic material, one further layer of the 25 $\mu$m silver foil was placed on top to act as a degrader. For the experiments performed at ISIS, ``fly-past'' mode was used such that $<$ 5 \% of the muons were implanted in the holder and all remaining muons that were not stopped in the sample were allowed to exit into a long tube and stop at the end, where their decay positrons were out of range of the detectors. This was not necessary at PSI, as the beam area was smaller than the well-centred sample. Data analysis was made using the program ``WiMDA'' \cite{Pratt3}, and ``Quantum'' \cite{Lord} was used to simulate the ALCs. For the simulations presented here, we used a muon-electron system without any nuclei, since the impact of nuclei were found to be negligible. The dynamical calculation with a Monte-Carlo (MC) powder averaging was used, with about 10,000 MC steps for each magnetic field. Further information on the analysis can be found in \cite{EPAPS}.

\begin{figure}
\includegraphics[scale=0.65]{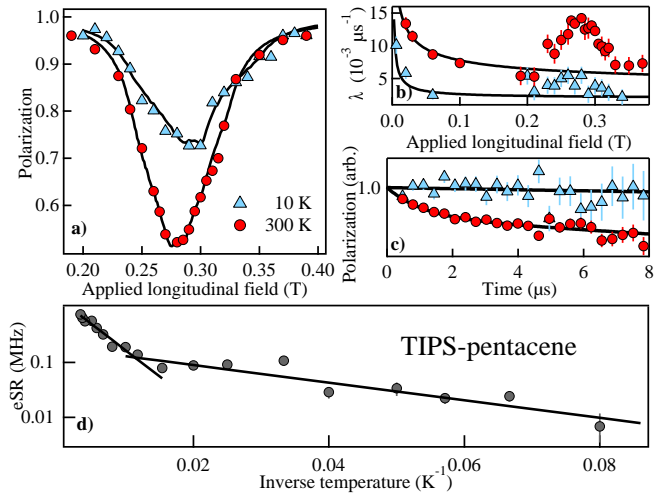}
\caption{(Color online) Muon data for TIPS-pentacene for T = 300 K (red circles) and T = 10 K (blue triangles). (a): The muon spin polarization around the avoided level crossings (ALC). Modeling for these ALCs is indicated by the black lines (see text). (b): The fitted field dependent exponential muon spin relaxation rate $\lambda$, showing there is a peak in the muon spin relaxation rate around the position of the ALCs. The lines show the expected power law dependence for the off-resonant relaxation rate. (c): Corresponding time dependent muon polarization at 0.27 T. Data are plotted using a linear scale and the black lines correspond to fits to an exponential function, from which the relaxation rate shown in (b) is extracted. (d): The eSR for TIPS-pentacene on an Arrhenius plot, with two characteristic energy scales (see text).}
\label{fig2}
\end{figure}

Figure \ref{fig2}a shows time-integrated muon polarization data near an ALC of polycrystalline TIPS-pentacene at high and low temperatures. The spectra exhibit ALC resonances close to 0.3 T corresponding to isotropic HFC constants of around 80 MHz. Further details of the ALC resonances can be found in \cite{EPAPS}. The one common and most prevalent feature in these spectra is the clear and unambiguous increase of the depth of the ALC minimum at high temperature. From the time-differential data and the field dependent relaxation rate, shown in Figures \ref{fig2}b-c, this increased ALC depth is clearly related to a higher relaxation rate of the muon spin. The relaxation rate begins to increase away from the expected power-law behavior (shown by the lines in Fig. \ref{fig2}b), with a pronounced peak centred on the location of the ALC. This discounts any non-relaxing phenomena, such as changes in muonium formation probability or background.

One possible explanation for the data shown in Figure \ref{fig2} could be a time-dependent modulation of the spin density sampled by the muon, as a result of periodic structural changes, such as phonons or intra-molecular vibrations. However, as is clear from the temperature independent width and position of the ALC, the isotropic and anisotropic HFCs are not particularly temperature dependent. One would expect a much larger temperature dependence of the HFCs, if this modulation was responsible. Any rotational molecular dynamics, such as those observed in C60 \cite{Roduner3}, can be discounted as our sample was solid without a rotational degree of freedom and there should nonetheless be an associated change in shape and/or position of the ALC. Finally, although surrounding nuclei spins can have an effect on the shape of the ALC, they cannot account for the dramatic increase in relaxation rate, because of their small coupling strength to the muon spin.

The only relaxation phenomenon that can reasonably describe the data shown in Figure \ref{fig2} is an electron spin relaxation rate (eSR) within the muon's experimental time window, where in the limit of intermediate eSR, (0.01 $<$ eSR $<$ $\sim$ 1 MHz) neither the position or absolute width of the ALC are significantly affected, whereas there is an increased time dependent relaxation \cite{Keitzman, Heming2}. The main effect of this relaxation is a change in ALC intensity, which is exactly as observed in the data shown in Figure \ref{fig2}a. Using a spin density matrix formalism \cite{Jurchescu, Lord} to model our spectra (solid lines in Figure \ref{fig2}a), we extracted an eSR of 0.76$\pm$0.03 MHz at 300 K and 0.02$\pm$0.01 MHz at 10 K. More importantly, we are also able to follow the eSR as a function of temperature from the amplitude increase in the ALC, thus helping to reveal the mechanism responsible for spin relaxation in organic materials. 

\begin{figure}
\includegraphics[scale=0.65]{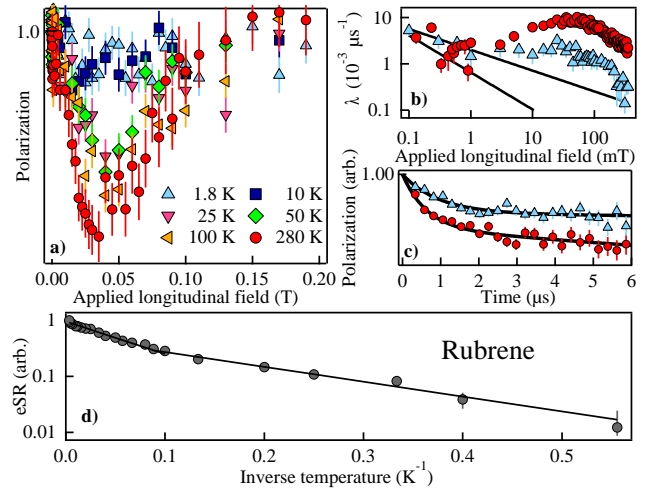}
\caption{(Color online) (a): Plot of the muons' field dependent spin polarization for rubrene, showing there is a broad and intense ALC.  (b): The fitted field dependent exponential relaxation rate $\lambda$. The lines show the expected power law dependency of the off-resonant background relaxation at high fields. (c): The corresponding time dependent muon spin polarization at 0.03 T for T = 1.8 K and T = 280 K, plotted using a linear scale. (d): The temperature dependence of the eSR shows Arrhenius behavior also with two characteristic energy scales.}
\label{fig3}
\end{figure}

Figure \ref{fig2}d shows the eSR for TIPS-pentacene plotted on a logarithmic scale against inverse temperature. On this Arrhenius plot, the eSR reduces linearly with two characteristic energy scales of 19$\pm$2 meV and 3.2$\pm$0.2 meV, which are a similar energy to the molecular vibrations common in organic materials, suggesting there is a coupling of eSR to the population of these vibrations. 

As discussed earlier, many of the theoretical models applied to explain eSR are transport based and as such, an Arrhenius behavior to eSR is not necessarily surprising. One would therefore expect the spin relaxation rate to be significantly determined by the mobility. In our case, where we measure an electron localized to a single molecule, one needs to discount a relaxation phenomenon resulting from any other free charges, present due to track electrons resulting from the muon's thermalization or doping from impurities. We show below that this is not the case and there is indeed a localized relaxation mechanism responsible for eSR. We compare the results from amorphous rubrene - a tetracene derivative (see Figure \ref{fig1}b) - with those from TIPS-pentacene. From Figures \ref{fig3}a-c it is immediately clear that there is a broad and intense ALC resonance at a low magnetic field, which is associated with a time differential relaxation. Furthermore, both the ALC amplitude and muon relaxation rate show a large magnitude change with temperature. In common with the TIPS-pentacene, the temperature dependence of eSR in rubrene shown in Figure \ref{fig3}d also has an Arrhenius nature, with two characteristic energy scales of 0.52$\pm$0.04 meV and 1.15$\pm$0.02 meV. Unfortunately it is not possible to model the rubrene data exactly as we have in TIPS-pentacene, since the local disorder results in an unquantifiable distribution of muon-electron HFCs. The temperature dependent eSR shown in Figure \ref{fig3}c is therefore in arbitrary units, but the trend is clearly the same as in TIPS-pentacene.

\begin{figure}
\includegraphics[scale=0.68]{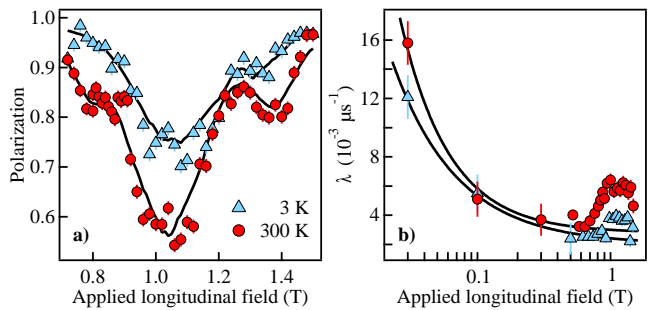}
\caption{(Color online) (a): A series of ALCs in polychrystalline Gaq3 around 1 T, for 300 K (red circles) and 3 K (blue triangles). (b): An increase in relaxation rate $\lambda$ is observed in the region where the ALCs are observed. The temperature dependence of both (a) and (b) are very similar to the acene molecules, suggesting a common mechanism.}
\label{fig4}
\end{figure}

The similarity between our TIPS-pentacene and rubrene data is quite surprising. As discussed earlier, the crystal structures of the materials are quite different, resulting in a high charge carrier mobility along 2 axes for TIPS-pentacene and no axes for amorphous rubrene. Given the similarities in results on these materials, it is clear that electron spin relaxation is not simply a transport phenomenon as is often assumed for organic materials \cite{Dediu1, Pramanik, Rybicki, Dediu2, Elliott, Dyakonov, Desai}.  In addition to any transport-based spin relaxation that may be present, there is an intra-molecular process giving spin relaxation. Interestingly, Arrhenius eSR is also observed in tetrathiafulvalene (TTF) \cite{Pratt2}, opening up the possibility that this local relaxation mechanism is more generally applicable across the range of molecular semiconductors. To assess the possibility of generality further, we performed high-field ALC measurements on tris(8-hydroxyquinolate)gallium (Gaq3), which has a tri-ligand molecular structure and does not contain an acene backbone (see Figure \ref{fig1}c). Gaq3 was synthesized using a published method \cite{Hernandez} and purified using train sublimation, resulting in a polycrystalline structure.  Approximately 800 mg of Gaq3 purified polycrystals was used in the muon experiments. In Figure \ref{fig4} we plot time-integrated muon spin polarization data for 300 K and 3 K, where a series of ALC resonances is found between 0.8 T and 1.5 T, around which an elevated muon spin relaxation rate is observed. Most importantly, the muon spin relaxation rate in the region of the ALC and its amplitude are significantly larger at 300 K compared to 3 K, with little or no changes to position or shape. This is very similar to all the previously presented data. Modeling in a similar manner to the other molecules, we find that the eSR rises from 0.02$\pm$0.01 MHz at 3 K to 0.85$\pm$0.02 MHz at 300 K. Whilst we do not have sufficient data to show an Arrhenius temperature dependence, nonetheless the similarity suggests that it is the same mechanism responsible for the eSR in all of the molecules presented here, in addition to TTF \cite{Pratt2}. We therefore suggest that the coupling of eSR to molecular vibrations could be a mechanism general to all small molecular systems. 

Interestingly, at the fields presented here (up to 1.5 T), any hyperfine coupling between protons and electrons should be quenched out. One possible mechanism would be the spin-orbit interaction, which is consistent with the small effect of deuteration on the intrinsic organic magnetoresistance \cite{Rolfe}. Previously, an Elliot-Yafet-like spin-orbit mechanism has been suggested as a possible driver of spin relaxation \cite{Cinchetti, Pramanik}, where the eSR is inversely proportional to the charge carrier mobility. Careful future muon studies of the effect of disorder or dopants on eSR may also reveal which transport related electron spin relaxation mechanism is relevant. Most importantly, however, the data presented here indicate that future theoretical models for spin transport in organic materials should also take into account relaxation when the charge carriers are residing on molecules, which could be very significant considering the diffusive nature of charge transport in these materials.

We would like to thank A. Stoykov and R. Scheuermann for the experimental support received at PSI and M. Somerton for the preparation of Gaq3. We would also like to thank E. Roduner, Z. Salman and R. Scheuermann for valued discussions with regards to the analysis and interpretation of our data. AJD acknowledges financial support from the EPSRC (grant EP/G054568/1) and the Leverhulme Trust. CB acknowledges financial support from the SNF (grant 200020-119784 and 200020-129484) and the NCCR program MaNEP. CAS acknowledges financial support from MaNEP whilst visiting the University of Fribourg.

\end{document}